\documentclass[runningheads]{llncs}
\usepackage[T1]{fontenc}
\usepackage{geometry}
\usepackage{amsmath}
\usepackage{graphicx}
\usepackage{enumerate}
\usepackage{booktabs} 
\usepackage{url} 
\usepackage{subcaption}
\usepackage{lipsum}
\usepackage{multirow}
\usepackage{xcolor}
\usepackage{lineno}
\usepackage{makecell}
\usepackage{algorithm}
\usepackage{algorithmic}
\usepackage{booktabs}
\usepackage{wasysym}
\usepackage{amssymb}
\usepackage{pifont}
\usepackage{caption}
\usepackage[export]{adjustbox}
\usepackage{hyperref} 
\usepackage{bm}
\usepackage{array}
\usepackage[figuresright]{rotating}
\usepackage[misc]{ifsym}
\usepackage{cite}
\usepackage{CJKutf8}
\usepackage{setspace}
\usepackage{capt-of}
\usepackage{placeins} 
\makeatletter
\def\UrlAlphabet{%
      \do\a\do\b\do\c\do\d\do\e\do\f\do\g\do\h\do\i\do\j%
      \do\k\do\l\do\m\do\n\do\o\do\p\do\q\do\r\do\s\do\t%
      \do\u\do\v\do\w\do\x\do\y\do\z\do\A\do\B\do\C\do\D%
      \do\E\do\F\do\G\do\H\do\I\do\J\do\K\do\L\do\M\do\N%
      \do\O\do\P\do\Q\do\R\do\S\do\T\do\U\do\V\do\W\do\X%
      \do\Y\do\Z}
\def\UrlDigits{\do\1\do\2\do\3\do\4\do\5\do\6\do\7\do\8\do\9\do\0}
\g@addto@macro{\UrlBreaks}{\UrlOrds}
\g@addto@macro{\UrlBreaks}{\UrlAlphabet}
\g@addto@macro{\UrlBreaks}{\UrlDigits}
\makeatother

\usepackage{tikz}
\newcommand*{\circled}[1]{\lower.7ex\hbox{\tikz\draw (0pt, 0pt)%
    circle (.5em) node {\makebox[1em][c]{\small #1}};}}

\begin{document}
\title{Enhancing Diagnostic Reliability of Foundation Model with Uncertainty Estimation in OCT Images}
\author{Yuanyuan Peng \inst{1\#}\and
Aidi Lin\inst{2\#}\and
Meng Wang\inst{3\#}\and
Tian Lin\inst{2}\and
Ke Zou\inst{4}\and
Yinglin Cheng\inst{2}\and
Tingkun Shi\inst{2}\and
Xulong Liao\inst{2}\and
Lixia Feng\inst{5}\and
Zhen Liang\inst{1}\and
Xinjian Chen\inst{6,7}\and
Huazhu Fu\inst{8 (\textrm{\Letter})}\and
Haoyu Chen\inst{2 (\textrm{\Letter})}
}
\titlerunning{FMUE}
\authorrunning{Y. Peng et al.}
%
\institute{
School of Biomedical Engineering, Anhui Medical University, Hefei, Anhui 230032, China \and
Joint Shantou International Eye Center, Shantou University and the Chinese University of Hong Kong, Shantou, Guangdong 515041, China.\and
Beth Israel Deaconess Medical Center, Harvard Medical School, 330 Brookline Ave, Boston, MA 02215, USA.\and
National Key Laboratory of Fundamental Science on Synthetic Vision and the College of Computer Science, Sichuan University, Chengdu, Sichuan 610065, China.\and
Department of Ophthalmology, First Affiliated Hospital of Anhui Medical University, Hefei, Anhui, China.\and
School of Electronics and Information Engineering, Soochow University, Suzhou, Jiangsu 215006, China.\and
State Key Laboratory of Radiation Medicine and Protection, Soochow University, Suzhou, Suzhou 215006, China.\and
Institute of High Performance Computing (IHPC), Agency for Science, Technology and Research (A*STAR), 1 Fusionopolis Way, \#16-16 Connexis, Singapore 138632, Republic of Singapore.
\\
\# Y. Peng, A. Lin, and M. Wang are the co-first authors.\\
\textrm{\Letter} Corresponding author: Huazhu Fu (\email{hzfu@ieee.org}), Haoyu Chen(\email{chy@jsiec.org}).
}
\maketitle              
\begin{abstract}
Inability to express the confidence level and detect unseen classes has limited the clinical implementation of artificial intelligence in the real-world. We developed a foundation model with uncertainty estimation (FMUE) to detect 11 retinal conditions on optical coherence tomography (OCT). In the internal test set, FMUE achieved a higher F1 score of 96.76\% than two state-of-the-art algorithms, RETFound and UIOS, and got further improvement with thresholding strategy to 98.44\%. In the external test sets obtained from other OCT devices, FMUE achieved an accuracy of 88.75\% and 92.73\% before and after thresholding. Our model is superior to two ophthalmologists with a higher F1 score (95.17\% vs. 61.93\% \&71.72\%). Besides, our model correctly predicts high uncertainty scores for samples with ambiguous features, of non-target-category diseases, or with low-quality to prompt manual checks and prevent misdiagnosis. FMUE provides a trustworthy method for automatic retinal anomalies detection in the real-world clinical open set environment.  

\end{abstract}
\newpage
\section{Introduction}
Retinal diseases are important causes of irreversible blindness and social burden~\cite{Ref1}. Screening of the susceptible population, early diagnosis, and timely management effectively reduce visual impairment and blindness. However, diagnosing retinal diseases requires well-trained ophthalmologists, whose shortage is unable to cope with the growing number of retinal disease patients. Furthermore, the distribution of medical resources is not even. In rural and underdeveloped regions, the shortage of ophthalmologists is worse. Optical coherence tomography (OCT) is a non-invasive, fast, high-resolution imaging technology that can visualize the cross-sectional retinal structure~\cite{Ref2,Ref3}. It is the gold standard for diagnosing most retinal diseases~\cite{Ref3,Ref4,Ref5}. However, interpreting OCT images is time-consuming and requires profound expertise. Therefore, developing an automatic retinal disease detection model based on OCT images has the promise of assisting clinical decision-making, reducing the workload of ophthalmologists, and facilitating blindness prevention.

Deep learning (DL) has been applied in retinal imaging, including OCT, to facilitate automatic diagnosis. In 2018, transfer learning was introduced to classify OCT images into normal and three diseases~\cite{Ref6}. A deep learning model was also developed to classify OCT volumes to different referral suggestions and diagnosis probabilities~\cite{Ref7}. Recently, RETFound, a foundation model pretrained on large-scale color fundus photography (CFP) and OCT images, has demonstrated great potential in detecting retinal diseases after fine-tuning~\cite{Ref8}. However, a significant downside of the standard AI model is that it only gives the prediction results without any information reflecting the reliability of the prediction, which may lead to low credibility of the model in real clinical implementation. Furthermore, the models are developed with a limited number of disease categories, and would counter unseen diseases (out of distribution, OOD) in real-world implementation and make incorrect predictions. The mistake of the model may lead to misdiagnosis or missed diagnosis and finally affect the prognosis of patients. 

In our previous study, we developed an uncertainty-inspired open-set learning (UIOS) model for CFP classification~\cite{Ref9}, which can output an uncertainty score in addition to the probabilities of disease categories. When the model encounters OOD data, such as unseen diseases, it will output a high uncertainty score above the threshold, indicating a double check by an ophthalmologist and preventing misdiagnosis. In the current study, we integrated a fine-tuned foundation model with uncertainty estimation (FMUE) in OCT images, enabling the capability of expressing the level of confidence and reliability of disease classification in open-set clinical implementation. Fig.~\ref{Overview} shows the training and inference process of our proposed FMUE framework. We compared our model performance with two recent state-of-the-art DL algorithms and two ophthalmologists to validate its effectiveness.

\section{Results}
\subsection{Performance on the internal test set }
In the internal test set with 19,046 images, our FMUE achieved an average of 96.76\% for F1 score (92.21\% to 100\%, Table 1), 96.66\% for accuracy (Table~2), 97.39\% for sensitivity (91.26\% to 100\%, Supplementary Table 1), 96.30\% for precision (83.09\% to 100\%, Supplementary Table 2), and 99.63\% for the area under the curve (AUC) of the receiver operating characteristic curve (ROC) (98.41\% to 100\%, Supplementary Fig. 1). The confusion metrics were shown in Supplementary Fig. 2. The performance of FMUE was superior to RETFound and UIOS, for example, the average F1 score of FMUE (96.76\%) was higher than that of RETFound (95.44\%, p = 0.0727, Supplementary Table 3) and UIOS (94.01\%, p=0.0386, Supplementary Table 3). 

We also used a threshold strategy to remove samples with high uncertainty. There were 9.92\% samples with uncertainty scores above the threshold in FMUE, which was lower than that in UIOS (12.21\%, Table~2 and Supplementary Table 4) and the distribution of uncertainty scores in the internal test set was similar to the validation set (Fig.~\ref{Fig2} (a)). The samples with high uncertainty score were 19.104-fold (95\% CI: 16.120-22.627, p<0.001) risk of being misclassified if they were not removed (Supplemental Table 5). Furthermore, after thresholding the OOD samples with high uncertainty scores, the performance metrics (Table~1, Supplemental Table 1-2) and confusion matrix (Supplementary Fig. 2) of FMUE were further improved; for example, the average F1 score improved to 98.44\% (Table 1), which was better than that of UIOS with thresholding (96.11\%, Table~1 and Supplementary Table 3).
\begin{figure}[h]
 \begin{center}
  \includegraphics[width=0.8\textwidth, height=0.8\textheight, keepaspectratio]{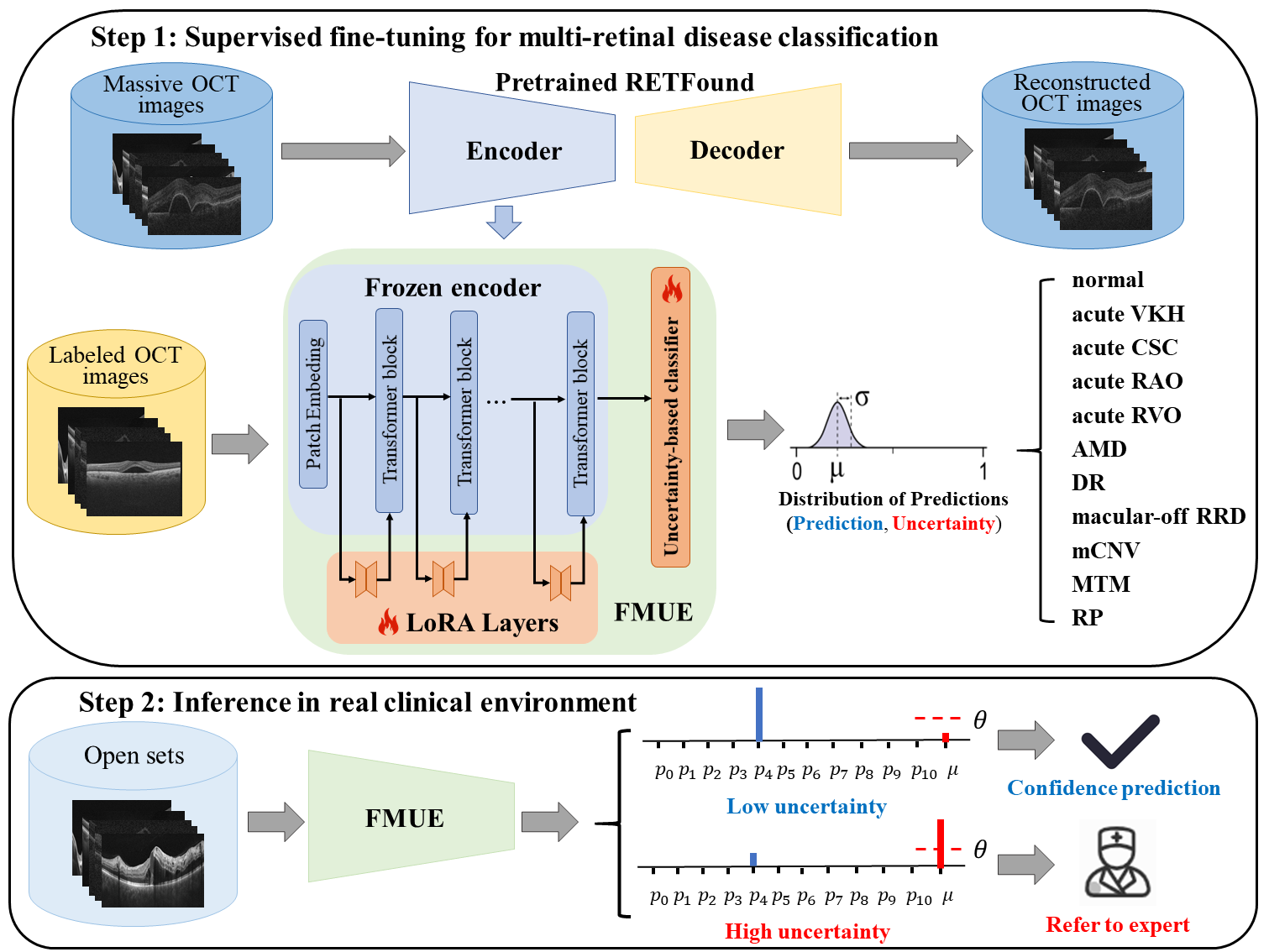}
 \end{center}
 \caption{\textbf{Schematic diagram of our FMUE for clinal work.} Step 1 adapted pretrained RETFound to multiple retinal disease classifications on OCT images by means of supervised fine-tuning on data with explicit label. We freeze the image encoder of RETFound (blue area) and insert additional trainable LoRA layers to RETFound for OCT image feature extraction. In addition, to increase the credibility of AI model prediction results, we developed an uncertainty-based classifier to obtain the final prediction result with corresponding uncertainty score. Step 2 shows the inference process of our FMUE in real clinical environment. When the model is fed with an image with obvious features of retinal disease in the training categories, our FMUE will give a diagnosis result with an uncertainty score below the threshold $\theta$ to indicate the diagnosis result is reliable. Conversely, when the input image contains ambiguous features or is OOD data, our model will give a high uncertainty score above the threshold $\theta$ to indicate the result is unreliable and refer the patient to an ophthalmologist for double-checking.}
 \label{Overview}
\end{figure}
\FloatBarrier
\vspace{-4cm}

\subsection{Performance on the external test sets}
The models were also tested on five external testing sets, which were publicly available, with various types of diseases and scanned using different models of OCT instruments (Supplemental Table 12). Overall, in a total of 7,416 images, the accuracy of FMUE (88.75\%) was higher than that of RETFound (88.08\%) and UIOS (80.66\%) (Table~2). Furthermore, after thresholding, the accuracy of FMUE and UIOS improved further. It was also noted that more samples were identified as having high uncertainty in the external test sets than in the internal test set (Fig.~\ref{Fig2} (a), Table~2 and Supplementary Fig. 3). 

The accuracy of FMUE after thresholding (92.73\%) was superior to UIOS after thresholding (87.60\%) (Table~2). Since the excluded samples differed between UIOS and FMUE, it is unfair to compare the performance of the two models after thresholding directly. We analyzed the accuracy by excluding different numbers of high uncertainty samples and found that the AUCs of the accuracy vs. percentage of samples were all higher in FMUE (97.03\%) than UIOS (89.23\%) (Fig.~\ref{Fig2}). Furthermore, the curves also showed that excluding the samples with high uncertainty improved the accuracy of classification.
\subsection{Out-of-distribution anomaly detection}
We evaluated the OOD anomaly detection performance of our FMUE using three OCT datasets with abnormal samples of non-target categories (NTC). FMUE detected 84.72\%, 89.55\%, and 91.17\% of samples with high uncertainty scores on NTC-internal, NTC-external and Low-Quality OCT datasets, respectively (Fig.~\ref{Fig2} (a)), which were more than those of UIOS (81.60\%, 86.39\%, and 89.91\% respectively). Furthermore, the distribution of uncertainty scores in the three OOD datasets is more skewed toward higher values in FMUE than UIOS.
\begin{figure}[h]
  \centering
  \includegraphics[width=1\textwidth, height=0.7\textheight, keepaspectratio]{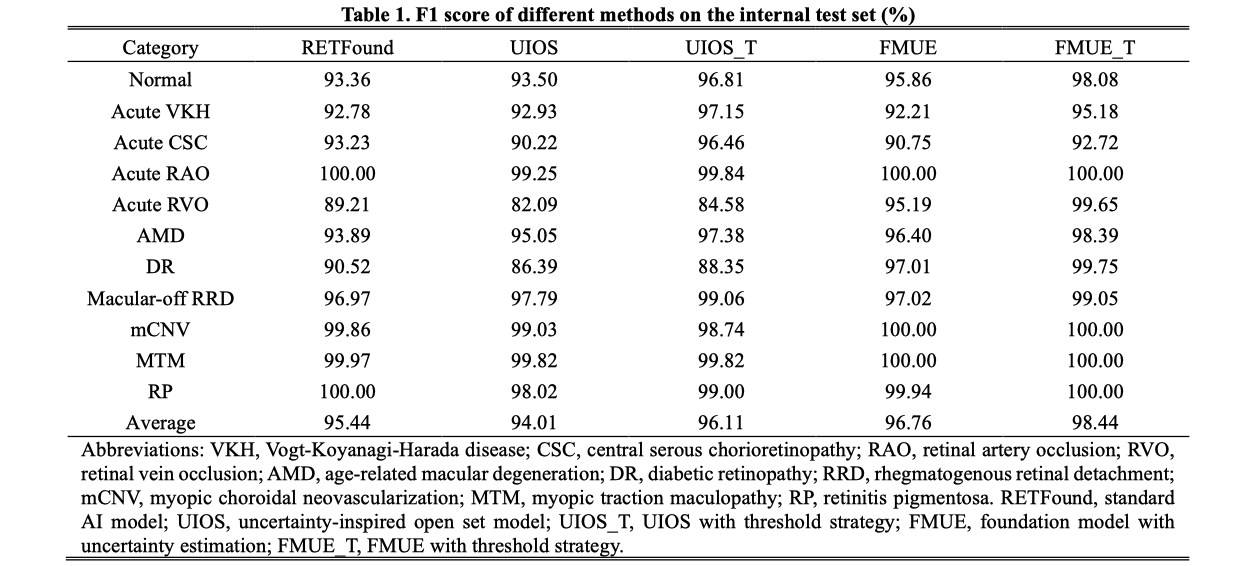}
  \label{Table1}
\end{figure}
\FloatBarrier
\subsection{Performance in human-model comparison (HMC)}
The human-model performance comparison results on the HMC set (Supplementary Table 4) are displayed in Table~3 and Supplementary Table 6-7. The average F1 scores of junior and senior doctors were 61.93\% and 76.72\%, respectively, which were significantly lower than that achieved by FMUE (95.17\%, p=0.0004 and 0.0039, respectively, Supplementary Table 3). After filtering out the samples with high uncertain scores in FMUE (14.4\%) and the samples with inconsistent results in two doctors (42.5\%), the F1 score of FMUE and doctors improved to 97.13\% and 83.67\%, respectively. FMUE still outperformed doctors, although the p-value did not reach statistical significance (p=0.0920, Supplementary Table 3). 

In the HMC set, our FMUE obtained high uncertainty scores in 144 images (14.4\%, Supplementary Table 8), and 54 images were misclassified without thresholding (5.4\%, Supplementary Table 8). The model uncertainty was positively associated with the misclassification made by FMUE on HMC set (OR=8.382, p<0.001), but not with the uncertainty of doctors (p=0.652) (Supplemental Table 9). The high uncertainty and misclassification of FMUE in the HMC set were primarily due to the similar features sharing in different diseases; for example, macular edema in diabetic retinopathy (DR) and retinal vein occlusion (RVO), or subtle features such as very small drusen that may be ignored and misclassified as normal conditional (Supplemental Fig. 4 and Supplemental Table 8).
\subsection{Examples and vision interpretation}

Fig.~\ref{Fig3} displays the heatmaps generated by Gradient-weighted Class Activation Mapping (Grad-CAM) to provide visual explanations for decisions made by our FMUE. Fig.~\ref{Fig3} a and Fig.~\ref{Fig3} b were two examples with typical features highlighted in red color and correctly predicted by FMUE with low uncertainty scores. RETFound and UIOS also made correct predictions, although UIOS outputted a high uncertainty score for the first image. 

Fig.~\ref{Fig3} c and Fig.~\ref{Fig3} d were two examples of target categories but with ambiguous features; Fig.~\ref{Fig3} e and Fig.~\ref{Fig3} f were two NTC examples. The model did not identify the features of these four images, as shown in the heatmaps. RETFound made an incorrect prediction without warning of the unreliability. UIOS and FMUE outputted high uncertainty scores to indicate unreliable classifications and a double check by ophthalmologists was needed.

\section{Discussion}
In the current study, we fine-tuned a foundation model and integrated uncertainty estimation for the task of retinal OCT multi-classification and OOD detection. The results showed that FMUE achieved better performance than RETFound and UIOS in multi-classification in both internal and external test datasets. FMUE also outperformed UIOS in OOD detection, which was absent in RETFound. In human-model comparison, FMUE had a higher F1 score than junior and senior doctors. The images with high uncertainty scores had a higher risk of misclassification.

The estimation of uncertainty is reliable and efficient in our model. Logistic regression analysis showed that the samples with uncertainty above the threshold had a 19.104 and 8.32-fold higher risk of misclassification in the internal test set and HMC set, respectively, suggesting the reliability of our FMUE. Furthermore, the curve of accuracy vs. percentage of samples showed that excluding the samples with high uncertainty improved the accuracy of classification (Fig 2). It supports the efficiency of uncertainty estimation in disease detection. In our method, the uncertainty score and prediction were optimized simultaneously, which may explain the reliability and efficiency of uncertainty estimation.

\begin{figure}[h]
 \begin{center}
  \includegraphics[width=0.85\textwidth, height=0.85\textheight, keepaspectratio]{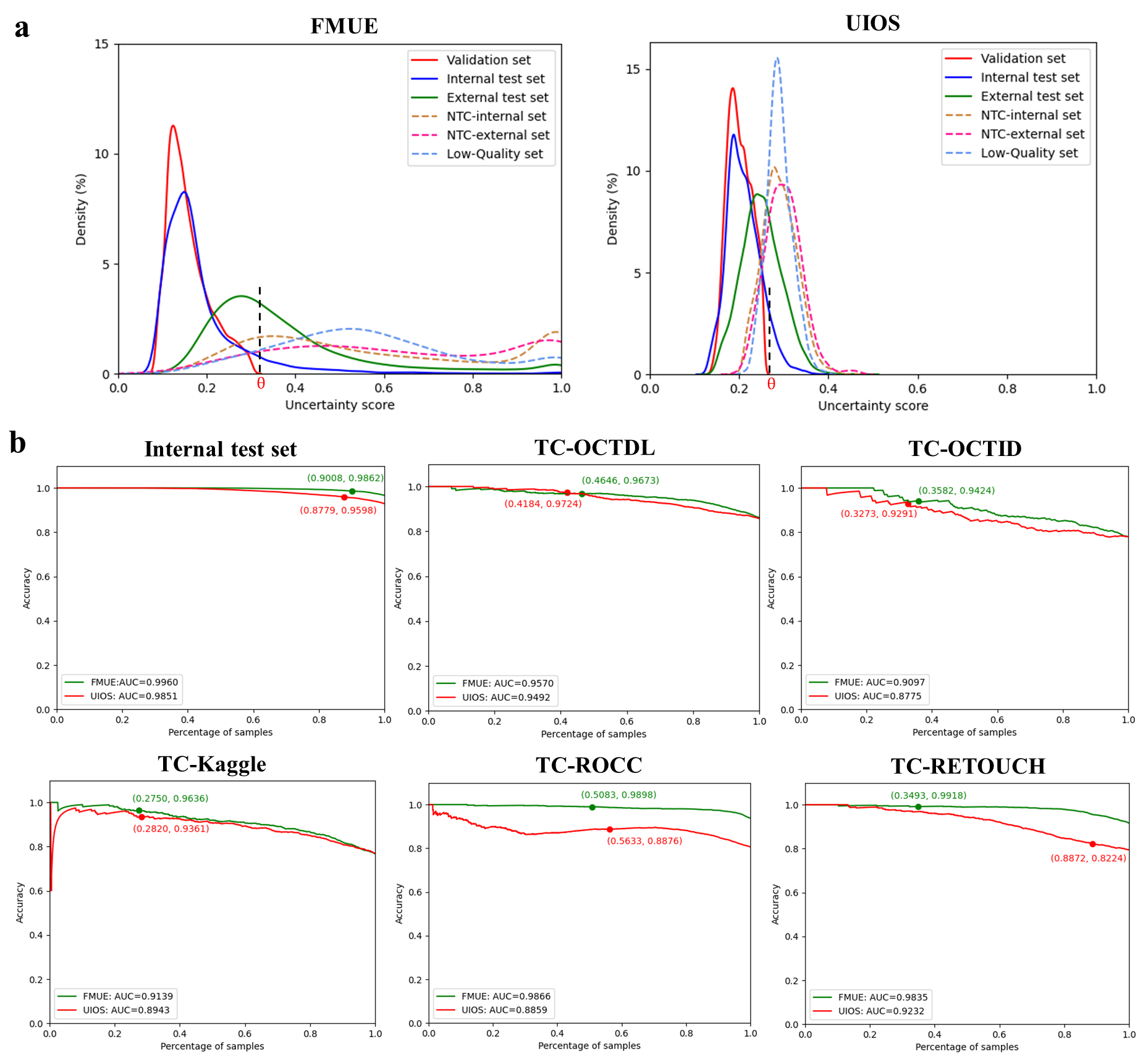}
 \end{center}
 \caption{\textbf{The performance of FMUE and UIOS on different datasets.} a. Uncertainty density distribution for different datasets in FMUE and UIOS. Solid lines indicate validation and test datasets for target categories of retinal diseases, while different colored dashed lines indicate different out-of-distribution datasets. $\theta$: threshold theta. b. The accuracy of FMUE and UIOS with different percentages of samples remained after excluding the high uncertainty samples on the internal test set and 5 external test sets. The green and red lines curve represent UIOS and FMUE, respectively. The dots on the curves indicate the coordinators of the threshold.}
 \label{Fig2}
\end{figure}
\FloatBarrier

The results show that FMUE is superior to RETFound and UIOS in target disease classification and OOD detection on OCT images. In comparison with RETFound, we integrated uncertainty estimation and had the capability to detect samples with ambiguous features and OOD samples unseen during training. Compared with UIOS, we used a transformer-based foundation model instead of a convolutional neural network as the backbone model, and fine-tuned it with LoRA, which kept the pretrained weights of the backbone network frozen. FMUE may benefit from the powerful feature extraction capability of the Vit-large/16 model pretrained on OCT images, even if only a small portion of weights were updated. 

\begin{figure}[h]
 \begin{center}
  \includegraphics[width=1\textwidth, height=1\textheight, keepaspectratio]{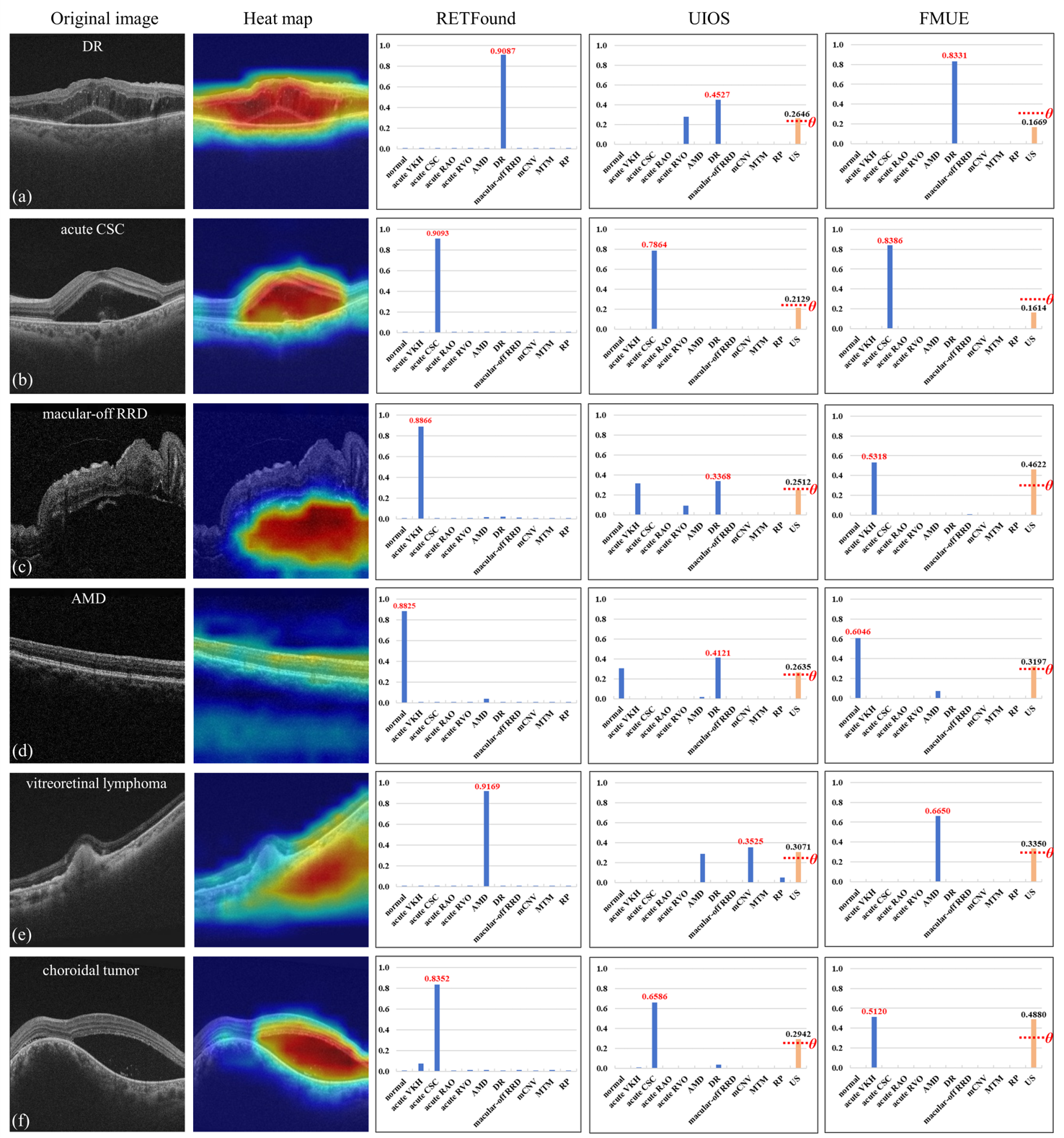}
 \end{center}
 \caption{\textbf{The visualization results of FMUE by Grad-CAM and the detection results of six samples of OCT images with RETFound, UIOS and our FMUE.} (a) and (b) are the samples with typical features of target diseases; (c) and (d) are the samples with ambiguous features of target diseases; (d) and (f) are OOD samples that are not included in the training category. Unlike RETFound, UIOS and FMUE provide prediction results and the corresponding uncertainty score to reflect the reliability of the prediction results. $\theta$ threshold theta.}
 \label{Fig3}
\end{figure}

Our FMUE showed better performance compared to both junior and senior ophthalmologists. In clinical practice, ophthalmologists usually interpret retinal diseases using 3D volume OCT images~\cite{Ref2}. However, evaluating 3D OCT images requires ophthalmologists to be highly focused and is a very time-consuming process. While making the prediction results based on 2D OCT image, the low diagnostic accuracy was made by doctors. Moreover, the inconsistency rates were higher than the high uncertainty rate in FMUE. This may be explained by the subjectivity of doctors (Supplementary Table 4). The inconsistency made by the two doctors is naturally present and belongs to the data uncertainty~\cite{Ref14}, which differs from the model uncertainty and explains why there is no relation for the uncertainty between our model and doctors (Supplementary Table 9). Furthermore, our FMUE model also achieved fast speed and provided visual interpretation of pathologic features (Fig.~\ref{Fig3}).

\FloatBarrier

\begin{figure}[h]
  \centering
  \includegraphics[width=1\textwidth, height=0.7\textheight, keepaspectratio]{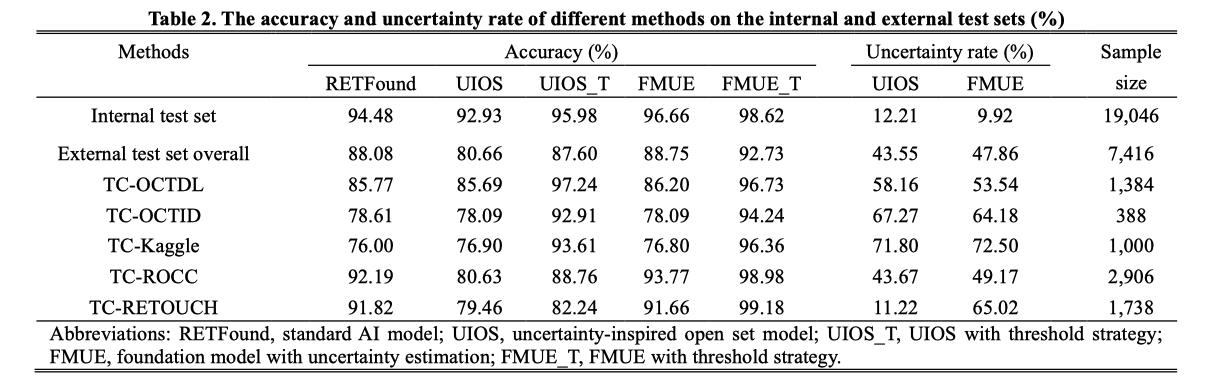}
\label{Table2}
\end{figure}
\vspace{-1.5cm}

\begin{figure}[h]
  \centering
  \includegraphics[width=1\textwidth, height=0.7\textheight, keepaspectratio]{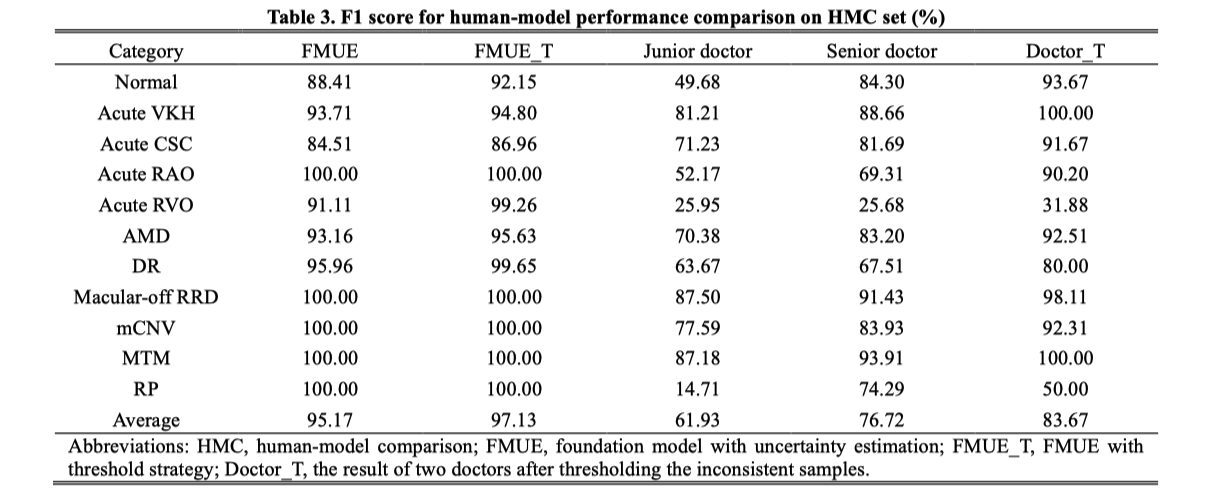}
  \label{Table3}
\end{figure}
\vspace{-1cm}

There are a few studies evaluating uncertainty for AI models based on OCT images. However, most did not explore how to detect OOD samples based on uncertainty~\cite{Ref10,Ref11,Ref12}. Seebock et al. trained a model with health OCT images and applied it in anomaly detection based on epistemic uncertainty, but this model cannot differentiate different diseases~\cite{Ref10}. Araújo et al. trained an efficient-Net V2-B0 using AMD staging images and used Dirichlet uncertainty estimation methods to detect near (DME, RVO, Stargardt) and far OOD (CFP), but only achieved low AUC with this model~\cite{Ref13}. In the current study, our model is trained on multiple common retinal disease images and capable of both classifications of 11 common conditions, and detection of uncommon diseases unseen during training (NTC) as OOD data using an uncertainty thresholding strategy. Therefore, the setting of our study is more applicable to clinical scenarios. 

There exist some limitations in this study. Firstly, although our FMUE model can achieve relatively accurate predictions for various retinal diseases, there are still 7.83\% and 40.40\% of the samples in the internal and external test sets exhibiting correct predictions with higher uncertainty than the threshold, requiring manual double-checking (Fig.~\ref{Fig2} ). It may be due to the instrument domain gap between the internal and external test set. In the next step, we will train more data obtained from different instruments except for Topcon to reduce the need for manual reconfirmation in these devices. Secondly, we only investigated the single-label classification, ignoring the issue of other coexisting diseases in the same OCT image. In the next stage, we will collect more multi-label classification data and explore uncertainty estimation methods to achieve reliable multi-label retinal disease detection. Thirdly, our FMUE only used single-mode OCT images and did not consider multi-modality imaging and valuable clinical text data. Therefore, multi-modality learning based on uncertainty estimation should be explored in further investigation.

In conclusion, our FMUE combined with threshold strategy can not only provide reliable diagnostic results for 11 types of retinal diseases and conditions, but also detect OOD samples that were not included during training, providing an automatic and trustworthy method for retinal anomaly detection using OCT images in real-world clinical scenario.

\section{Methods}
\subsection{Target categories OCT datasets}
This study adhered to the tenets of the Declaration of Helsinki and was approved by the Institutional Board of Joint Shantou International Eye Center of Shantou University and the Chinese University of Hong Kong (JSIEC). All the images were deidentified and encrypted to protect the security and privacy of personal information, and the informed consent from patients was waived. All the images were centered in the macula and collected from Joint Shantou International Eye Center using the electronic medical record. In addition, the images scanned at different times during the follow-up were also included.

The OCT retinal images of target categories were obtained from two OCT devices: Triton DRI OCT (Topcon, Tokyo, Japan) and 3D OCT‐2000 (Topcon Corporation, Tokyo, Japan). These datasets included 11 relatively common diseases and conditions (Supplementary Table 10): normal, acute Vogt-Koyanagi-Harada (VKH) disease, acute central serous chorioretinopathy (CSC), acute retinal artery occlusion (RAO), acute RVO, age-related macular degeneration (AMD) including dry and wet subtypes, DR, macular-off rhegmatogenous retinal detachment (RRD), myopic choroidal neovascularization (mCNV), myopic tractional maculopathy (MTM) and retinitis pigmentosa (RP). 

The inclusion criteria for these diseases/conditions are listed in Supplementary Table 10. All retinal OCT images were labeled with 11 diseases or conditions according to the characteristic features. In the first round, the masked images were sent to a junior grader (T.L.) to include those images with characteristic OCT features. The images with poor image quality affecting the image analysis and cases with uncertain diagnosis or comorbidity with other retinal diseases were also excluded. In the second round, two senior graders (A.L. and X.L.) were trained to label the images. They achieved a high agreement with one expert ($Kappa\geq 0.8$) on 100 images randomly selected from the dataset. After certification, the graders labeled the images independently. In case of disagreement, an experienced retinal expert (H.C.) made the final decision (Supplementary Fig. 5).

After two rounds of annotation, a total of 102,224 OCT images from 881 eyes of 784 subjects were collected. Based on the patient-based split policy, the images of each disease/condition were randomly split into the training, validation and test set in the ratio of 6:2:2. The numbers of images in each category within each dataset are listed in Supplementary Table 11. 

To further evaluate the generalization ability of our FMUE in detecting retinal diseases, we also conducted experiments on five public datasets obtained from various OCT instruments~\cite{Ref6,Ref15,Ref16,Ref17}~\footnote{https://rocc.grand-challenge.org}. We only included target category samples with characteristic features from the original datasets named TC-OCTDL, TC-OCTID, TC-Kaggle, TC-ROCC, and TC-RETOUCH, respectively. It must be noted that several types of OCT devices were used to obtain these images, including RTVue XR, Cirrus, Spetralis, 3D OCT-1000 and 3D OCT-2000. The numbers of images in each category within each dataset are listed in Supplementary Table 12.
\subsection{OOD datasets}
We used two NTC retinal disease datasets and a low-quality OCT image dataset to investigate the ability of FMUE in detecting retinal abnormalities outside the categories of the training set. The first was 6,656 OCT images collected from our clinic with retinal diseases outside the categories of the training set, which were obtained from Triton and 3D OCT‐2000 OCT devices in our clinic and called the NTC-internal dataset. The second included 214 images of vitreoretinal lymphoma collected from three foreign institutes, 181 images of epiretinal membrane and vitreomacular interface disease from the OCTDL dataset, and 102 images of macular hole from the OCTID dataset, which were called the NTC-external dataset. It should be noted that the OCT images in the NTC-external dataset were scanned using various types of OCT devices, including RTVue XR, Spectralis, and Cirrus. The low-quality OCT dataset was obtained from Triton OCT device in our clinic. It comprised 793 indistinguishable OCT images, primarily due to severe media opacity, image artifacts, or resolution reduction. The detailed information of NTC-internal and NTC-external is listed in Supplementary Table 13.
\subsection{Model development}
Fig.~\ref{Overview}  shows the training and inference process of our proposed FMUE framework. In the training stage, we discarded the decoder of RETFound~\cite{Ref8} and took its encoder as our backbone network to extract the high-level feature information contained in OCT images, followed by an uncertainty-based classifier to obtain the final prediction result with corresponding uncertainty score, which was different from the standard AI model only assigning a probability value to each category of retinal disease included in the training set and taking the category with the highest probability value as the final prediction result without any information reflecting the reliability of the final decision. If the standard AI model made incorrect predictions without any risk information prompts, it may bring serious consequences to clinical practice, especially in open-set clinical implementation. Uncertainty estimation can enable the capability of expressing the level of confidence and increase the credibility of AI model prediction results in open-set clinical implementation. Similar to our previous work~\cite{Ref9}, an uncertainty-based classifier was achieved by evidential and Dirichlet distribution-based subjective logistic uncertainty theory, which was described in the supplementary materials. In addition, to effectively adapt the pretrained backbone network to our retinal disease classification task, we introduced a simple and effective adaption strategy, Low-Rank Adaption (LoRA), for model optimization~\cite{Ref19}. Different from the fully fine-tuning training strategy~\cite{Ref8,Ref20,Ref21,Ref22}, the LoRA-based adaption strategy kept the pretrained weights of the backbone network frozen and automatically adjusted the weights between layers in the backbone network to improve model performance and reduce memory consumption and training time. A detailed introduction to the LoRA-based optimization strategy in this study can be found in the supplementary materials.

After model training, we can use fine-tuned FMUE for real clinical practice work, which can generate the final prediction result with an uncertainty score as shown in the second step of Fig.~\ref{Overview} . Different from the standard AI classification model, our FMUE can not only provide the final diagnosis result but also obtain an uncertainty score to indicate the reliability of the diagnosis result in inference stage. If the uncertainty score is higher than the threshold, a double-check by an experienced grader or ophthalmologist is required. In this scheme, although lower uncertainty scores indicate higher credibility, a larger number of OCT images are also excluded. Therefore, the optimal threshold point must strike an appropriate balance between prediction accuracy, certainty, the number of OCT images excluded due to uncertainty, and the incidence of retinal disease in the remaining OCT images. To obtain such optimal threshold, the OCT images with the highest uncertainty score were excluded one by one. Meanwhile, the accuracy and incidence of retinal disease in the remaining OCT images in the validation dataset were recalculated. Once one of the indicators decreases from the initial testing level, the process will stop.

\subsection{Human-model comparison}
The comparison was carried out between our FMUE and two ophthalmologists from JSIEC, including a junior doctor (Y.C.) and a retinal specialist (T.S.) with clinical experiences of 5 and 10 years, respectively. A total of 1,000 OCT images were selected randomly from the internal test set, namely HMC set. The number of each category within each dataset is listed in Supplementary Table 4. We analyzed their diagnostic ability for retinal diseases and the uncertainty in diagnosing retinal diseases, respectively. It is worth noting that the uncertainty of doctors was defined as their inconsistency in HMC set annotation.

\subsection{Interpretation}
DL models are often referred to as "black box" entities and lack interpretation of the results~\cite{Ref23,Ref24}. To improve transparency and interpretability, we applied the Grad-CAM technique to aid the interpretation of the results, which can capture the regions in the image that are relevant to the final classification result by calculating the gradient of a certain layer in the deep neural network~\cite{Ref25}.

\subsection{Statistical analysis}
To comprehensively and fairly evaluate the classification performance of different methods, accuracy, precision, sensitivity and F1 score were used. In addition, AUC was calculated using the open-source package scikit-learn (version 1.0.2). The factors associated with the model uncertainty were investigated using univariate logistic regression analysis. Associations were presented as odds ratio (OR) with a 95\% confidence interval (CI). Statistical analyses were performed using SPSS software version 19 (SPSS, Inc., Chicago, IL, USA).

\section{Code Availability}
All codes are available at \url{https://github.com/yuanyuanpeng0129/FMUE}.
\section{Data Availability}
Data from OCTDL is available at \url{https://ieee-dataport.org/documents/octdl-optical-coherence-tomography-dataset-image-based-deep-learning-methods}.

\noindent Data from OCTID is available at \url{https://dataverse.scholarsportal.info/dataverse/OCTID}.

\noindent Data from Kaggle is available at \url{https://doi.org/10.17632/rscbjbr9sj.3}.

\noindent Data from ROCC is available at \url{https://rocc.grand-challenge.org}.

\noindent Data from RETORCH is available at \url{https://retouch.grand-challenge.org}.

\noindent Additional data sets supporting the findings of this study were not publicly available due to the confidentiality policy of the Chinese National Health Council and institutional patient privacy regulations. However, they were available from the corresponding authors upon request. For replication of the findings and/or further academic and AI-related research activities, data may be requested from corresponding author H. Chen within 10 working days. Source data are provided in this paper.

\section{Acknowledgements}
This research is supported by the National Key R\&D Program of China (2018 YFA0701700 to H. C. and X. C.), Agency for Science, Technology and Research (A*STAR) Career Development Fund (C222812010 to H. F.), Central Research Fund (“Robust and Trustworthy AI system for Multi-modality Healthcare” to H. F.), the National Nature Science Foundation of China (U20A20170 to X. C.), Shantou Science and Technology Program (190917085269835 to H. C.), and 2020 Li Ka Shing Foundation Cross-Disciplinary Research Grant (2020LKSFG14B to H. C.).

\bibliographystyle{IEEEtran}
\bibliography{references}

\begin{thebibliography}{10}
\providecommand{\url}[1]{#1}
\csname url@samestyle\endcsname
\providecommand{\newblock}{\relax}
\providecommand{\bibinfo}[2]{#2}
\providecommand{\BIBentrySTDinterwordspacing}{\spaceskip=0pt\relax}
\providecommand{\BIBentryALTinterwordstretchfactor}{4}
\providecommand{\BIBentryALTinterwordspacing}{\spaceskip=\fontdimen2\font plus
\BIBentryALTinterwordstretchfactor\fontdimen3\font minus \fontdimen4\font\relax}
\providecommand{\BIBforeignlanguage}[2]{{%
\expandafter\ifx\csname l@#1\endcsname\relax
\typeout{** WARNING: IEEEtran.bst: No hyphenation pattern has been}%
\typeout{** loaded for the language `#1'. Using the pattern for}%
\typeout{** the default language instead.}%
\else
\language=\csname l@#1\endcsname
\fi
#2}}
\providecommand{\BIBdecl}{\relax}
\BIBdecl

\bibitem{Ref1}
A.~Mariotti and D.~Pascolini, ``Global estimates of visual impairment,'' \emph{Br J Ophthalmol}, vol.~96, no.~5, pp. 614--8, 2012.

\bibitem{Ref2}
F.~Pichi, R.~Dolz-Marco, J.~H. Francis, A.~Au, J.~L. Davis, A.~Fawzi, S.~Gattousi, D.~A. Goldstein, P.~A. Keane, E.~Miserocchi \emph{et~al.}, ``Advanced oct analysis of biopsy-proven vitreoretinal lymphoma,'' \emph{American journal of ophthalmology}, vol. 238, pp. 16--26, 2022.

\bibitem{Ref3}
A.~Lin, X.~Mai, T.~Lin, Z.~Jiang, Z.~Wang, L.~Chen, and H.~Chen, ``Research trends and hotspots of retinal optical coherence tomography: A 31-year bibliometric analysis,'' \emph{Journal of Clinical Medicine}, vol.~11, no.~19, p. 5604, 2022.

\bibitem{Ref4}
R.~Chopra, S.~K. Wagner, and P.~A. Keane, ``Optical coherence tomography in the 2020s—outside the eye clinic,'' \emph{Eye}, vol.~35, no.~1, pp. 236--243, 2021.

\bibitem{Ref5}
A.~Lin, H.~Xia, A.~Zhang, X.~Liu, and H.~Chen, ``Vitreomacular interface disorders in proliferative diabetic retinopathy: an optical coherence tomography study,'' \emph{Journal of Clinical Medicine}, vol.~11, no.~12, p. 3266, 2022.

\bibitem{Ref6}
D.~S. Kermany, M.~Goldbaum, W.~Cai, C.~C. Valentim, H.~Liang, S.~L. Baxter, A.~McKeown, G.~Yang, X.~Wu, F.~Yan \emph{et~al.}, ``Identifying medical diagnoses and treatable diseases by image-based deep learning,'' \emph{cell}, vol. 172, no.~5, pp. 1122--1131, 2018.

\bibitem{Ref7}
J.~De~Fauw, J.~R. Ledsam, B.~Romera-Paredes, S.~Nikolov, N.~Tomasev, S.~Blackwell, H.~Askham, X.~Glorot, B.~O’Donoghue, D.~Visentin \emph{et~al.}, ``Clinically applicable deep learning for diagnosis and referral in retinal disease,'' \emph{Nature medicine}, vol.~24, no.~9, pp. 1342--1350, 2018.

\bibitem{Ref8}
Y.~Zhou, M.~A. Chia, S.~K. Wagner, M.~S. Ayhan, D.~J. Williamson, R.~R. Struyven, T.~Liu, M.~Xu, M.~G. Lozano, P.~Woodward-Court \emph{et~al.}, ``A foundation model for generalizable disease detection from retinal images,'' \emph{Nature}, vol. 622, no. 7981, pp. 156--163, 2023.

\bibitem{Ref9}
M.~Wang, T.~Lin, L.~Wang, A.~Lin, K.~Zou, X.~Xu, Y.~Zhou, Y.~Peng, Q.~Meng, Y.~Qian \emph{et~al.}, ``Uncertainty-inspired open set learning for retinal anomaly identification,'' \emph{Nature Communications}, vol.~14, no.~1, p. 6757, 2023.

\bibitem{Ref14}
K.~Zou, Z.~Chen, X.~Yuan, X.~Shen, M.~Wang, and H.~Fu, ``A review of uncertainty estimation and its application in medical imaging,'' \emph{Meta-Radiology}, p. 100003, 2023.

\bibitem{Ref10}
P.~Seeb{\"o}ck, J.~I. Orlando, T.~Schlegl, S.~M. Waldstein, H.~Bogunovi{\'c}, S.~Klimscha, G.~Langs, and U.~Schmidt-Erfurth, ``Exploiting epistemic uncertainty of anatomy segmentation for anomaly detection in retinal oct,'' \emph{IEEE transactions on medical imaging}, vol.~39, no.~1, pp. 87--98, 2019.

\bibitem{Ref11}
X.~Wang, F.~Tang, H.~Chen, L.~Luo, Z.~Tang, A.-R. Ran, C.~Y. Cheung, and P.-A. Heng, ``Ud-mil: uncertainty-driven deep multiple instance learning for oct image classification,'' \emph{IEEE journal of biomedical and health informatics}, vol.~24, no.~12, pp. 3431--3442, 2020.

\bibitem{Ref12}
X.~Liu, K.~Zhou, J.~Yao, M.~Wang, and Y.~Zhang, ``Contrastive uncertainty based biomarkers detection in retinal optical coherence tomography images,'' \emph{Physics in Medicine \& Biology}, vol.~67, no.~24, p. 245012, 2022.

\bibitem{Ref13}
T.~Ara{\'u}jo, G.~Aresta, U.~Schmidt-Erfurth, and H.~Bogunovi{\'c}, ``Few-shot out-of-distribution detection for automated screening in retinal oct images using deep learning,'' \emph{Scientific Reports}, vol.~13, no.~1, p. 16231, 2023.

\bibitem{Ref15}
M.~Kulyabin, A.~Zhdanov, A.~Nikiforova, A.~Stepichev, A.~Kuznetsova, M.~Ronkin, V.~Borisov, A.~Bogachev, S.~Korotkich, P.~A. Constable \emph{et~al.}, ``Octdl: Optical coherence tomography dataset for image-based deep learning methods,'' \emph{Scientific Data}, vol.~11, no.~1, p. 365, 2024.

\bibitem{Ref16}
P.~Gholami, P.~Roy, M.~K. Parthasarathy, and V.~Lakshminarayanan, ``Octid: Optical coherence tomography image database,'' \emph{Computers \& Electrical Engineering}, vol.~81, p. 106532, 2020.

\bibitem{Ref17}
H.~Bogunovi{\'c}, F.~Venhuizen, S.~Klimscha, S.~Apostolopoulos, A.~Bab-Hadiashar, U.~Bagci, M.~F. Beg, L.~Bekalo, Q.~Chen, C.~Ciller \emph{et~al.}, ``Retouch: The retinal oct fluid detection and segmentation benchmark and challenge,'' \emph{IEEE transactions on medical imaging}, vol.~38, no.~8, pp. 1858--1874, 2019.

\bibitem{Ref19}
E.~J. Hu, Y.~Shen, P.~Wallis, Z.~Allen-Zhu, Y.~Li, S.~Wang, L.~Wang, and W.~Chen, ``Lora: Low-rank adaptation of large language models,'' \emph{arXiv preprint arXiv:2106.09685}, 2021.

\bibitem{Ref20}
X.~Li, L.~Shen, M.~Shen, F.~Tan, and C.~S. Qiu, ``Deep learning based early stage diabetic retinopathy detection using optical coherence tomography,'' \emph{Neurocomputing}, vol. 369, pp. 134--144, 2019.

\bibitem{Ref21}
D.~B. Russakoff, S.~S. Mannil, J.~D. Oakley, A.~R. Ran, C.~Y. Cheung, S.~Dasari, M.~Riyazzuddin, S.~Nagaraj, H.~L. Rao, D.~Chang \emph{et~al.}, ``A 3d deep learning system for detecting referable glaucoma using full oct macular cube scans,'' \emph{Translational Vision Science \& Technology}, vol.~9, no.~2, pp. 12--12, 2020.

\bibitem{Ref22}
A.~Sunija, S.~Kar, S.~Gayathri, V.~P. Gopi, and P.~Palanisamy, ``Octnet: A lightweight cnn for retinal disease classification from optical coherence tomography images,'' \emph{Computer methods and programs in biomedicine}, vol. 200, p. 105877, 2021.

\bibitem{Ref23}
L.~Gao and L.~Guan, ``Interpretability of machine learning: Recent advances and future prospects,'' \emph{IEEE MultiMedia}, 2023.

\bibitem{Ref24}
S.~A. Martin, F.~J. Townend, F.~Barkhof, and J.~H. Cole, ``Interpretable machine learning for dementia: a systematic review,'' \emph{Alzheimer's \& Dementia}, vol.~19, no.~5, pp. 2135--2149, 2023.

\bibitem{Ref25}
R.~R. Selvaraju, M.~Cogswell, A.~Das, R.~Vedantam, D.~Parikh, and D.~Batra, ``Grad-cam: Visual explanations from deep networks via gradient-based localization,'' in \emph{Proceedings of the IEEE international conference on computer vision}, 2017, pp. 618--626.

\end{thebibliography}

\end{document}